\documentclass[journal,onecolumn,12pt,twoside]{IEEEtranTCOM}
\usepackage{graphicx,amsmath,epsfig,times}
\usepackage{psfrag,cite}
\usepackage{amstext,amssymb,latexsym}
\usepackage{setspace}
\usepackage{algorithmic, algorithm}
\DeclareMathOperator*{\argmax}{arg\,max}

\onecolumn
\begin{document}

\title{Practical Methods for Wireless Network Coding with Multiple Unicast Transmissions}
\author{Tu\u{g}can~Akta\c{s}, {A.~\"{O}zg\"{u}r~Y\i lmaz}, {Emre~Akta\c{s}}
\thanks{Tu\u{g}can~Akta\c{s} and {A.~\"{O}zg\"{u}r~Y\i lmaz} are with the Department of Electrical and Electronics Engineering, Middle East Technical University (e-mail: aoyilmaz@metu.edu.tr, taktas@metu.edu.tr) and {Emre~Akta\c{s}} is with the Department of Electrical and Electronics Engineering, Hacettepe University (e-mail: aktas@ee.hacettepe.edu.tr), Ankara, Turkey. This work was presented in part at the 2012 IEEE Wireless Communications and Networking Conference (WCNC 2012), Paris, France under the title "Practical Wireless Network Coding and Decoding Methods for Multiple Unicast Transmissions"}}

\maketitle
\begin{abstract}
We propose a simple yet effective wireless network coding and decoding technique for a multiple unicast network. It utilizes spatial diversity through cooperation between nodes which carry out distributed encoding operations dictated by generator matrices of linear block codes. In order to exemplify the technique, we make use of greedy codes over the binary field and show that the arbitrary diversity orders can be flexibly assigned to nodes. Furthermore, we present the optimal detection rule for the given model that accounts for intermediate node errors and suggest a low-complexity network decoder using the sum-product (SP) algorithm. The proposed SP detector exhibits near optimal performance. We also show asymptotic superiority of network coding over a method that utilizes the wireless channel in a repetitive manner without network coding (NC) and give related rate-diversity trade-off curves. Finally, we extend the given encoding method through selective encoding in order to obtain extra coding gains.
\end{abstract}


\section{Introduction}
In order to counteract the effects of fading in wireless communication networks, many ways of creating diversity for transmitted data have been proposed. Utilizing the spatial diversity inherent in wireless channels, cooperative communication \cite{Erkip03} has been of great interest in recent years. In \cite{Laneman04, Laneman06} three methods to be used by relay nodes are described: amplify-forward (AF), decode-forward (DF) and detect-forward (DetF). The AF method attains full diversity, whereas other two cannot, unless the propagation of errors resulting from the decoding/detection operation is avoided, for example by using a selective transmission strategy that advocates the forwarding of only reliable enough packets. One of the various ways to handle this problem is using CRC-based methods, which results in loss of spectral efficiency due to drop of a packet with only a few bit errors. An on/off weighting based on relay signal-to-noise power ratio (SNR) is given in \cite{Onat08}. Weighting of the signals either at the relay or at the receiver using the relay error probability is proposed in \cite{Wang07, Wang08}. Yet another idea is transmitting the log-likelihood ratios (LLR) of bits \cite{Yang07}. However, the soft information relaying methods in \cite{Wang07, Wang08, Yang07} suffer from quantization errors and high peak to average ratio problems. In addition, the AF method requires hardware modifications on modern-day communications systems and the DF method leads to high complexity decoding operations especially for the relays. As an alternative, relays may use the simple DetF method, which is shown to avoid error propagation in \cite{Laneman06}, if the error probabilities at relays are known and the maximum a posteriori probability (MAP) detection is employed at the receiver. 

NC was initially proposed to enhance network throughput in wired systems with error-free links of unit capacity\cite{Ahlswede00}. Later studies exhibited the good performance of random linear NC \cite{Ho06}. In wireless networks with nodes naturally overhearing transmissions, NC can be utilized to create diversity, reduce routing overhead, and introduce MAC layer gains as discussed for practical systems in \cite{Medard08}. Although most of the work in the literature concentrate on multicast transmission \cite{Renzo10}; we deal with a network involving multiple unicast transmissions, which is inherent in real-life scenarios. Hence we formulate a multiple unicast transmission problem such that for each unicast transmission, there is a distinct diversity that is to be improved via spatial opportunities.  

We consider a simple NC scheme based on DetF. Given a relay combining strategy, which we represent by a generator matrix and a vector of transmit schedule, we investigate the diversity order of each source, which can be unequal. We propose a novel method for designing the generator matrix based on linear block codes over the binary field. The proposed method is very flexible in that any set of desired diversity levels for the sources can be achieved with the highest NC rate possible. The diversity analysis relies on an optimal MAP decoder at the destination which employs the reliability information of the intermediate nodes and avoids loss of diversity due to error propagation \cite{Laneman06,Renzo11b}. The numerical complexity of the given optimal decoder can be impractical. Thus we propose a practical approximation of the MAP detector: the SP network decoder. 

A study based on flexible network codes in a two-source two-relay system with emphasis on unequal error protection is \cite{Renzo11}, where authors propose a suboptimal detection rule (distributed minimum distance detector) that is known to result in diversity order loss. Our scheme captures full diversity due to the use of the SP detector with intermediate node reliability information. In \cite{Xiao2006}, a multicast scenario is investigated (again under additive white Gaussian noise (AWGN) assumption) for obtaining an optimal energy allocation scheme in order to minimize bit error rate at the sink nodes. In \cite{Xiao08}, performance of a multiple hop network with no fading assumption is analyzed in terms of effects of the bit errors at the relays utilizing a technique known as error event enumeration.  Similarly in \cite{Xiao09}, optimal detection rule to be used at the destination node is presented for AWGN channels in addition to the description of a genie-aided decoder which yields a lower bound on the performance of the optimal detector. Different than \cite{Xiao08} and \cite{Xiao09}, we consider faded wireless links and give optimal detection rule corresponding to this realistic scenario. Futhermore, the operation at the intermediate nodes in our scheme is DetF as opposed to the more complex DF in \cite{Xiao08,Xiao09}. One of the studies closest to ours is \cite{Xiao10}, where the NC operation is fixed in construction yielding very large Galois field (GF) sizes for increasing network size and relay nodes carry out complicated DF operation for each transmission they overhear. Similarly in \cite{Wang2011}, DF is used in a fixed single-relay two-user scenario in order to provide diversity-multiplexing trade-off for NC. However, our results indicate that any diversity order can be achieved for any unicast transmission even with the GF of size 2 by using linear block codes as the network codes and simple DetF. Also independent from our work, in \cite{Renzo11b, Renzo11c}, results concerning diversity analysis for a system model resembling ours have been obtained. Similarly in a recent work \cite{Renzo11d}, additional coding gain analysis is given for a multi-source multi-relay network with relays having no data to be transmitted. Our model is generalized in the sense that each node both acts as a source node with its own data to be conveyed over the network and as an intermediate node serving as a means for combining and relaying others' data. Moreover, the  proposed detection rules of \cite{Renzo11b, Renzo11c, Renzo11d} result in exponential decoding complexity in the number of transmissions, since they are based on maximum likelihood sequence estimation. 
Recently in \cite{Lu2010}, a wireless broadcast network with block erasures is considered and a network coding scheme is proposed for retransmissions. The improvement in the number of retransmissions  for the downlink channel with respect to the conventional automatic-repeat-request mechanism is clearly shown. Another recent work \cite{Xiao11} identifies the diversity-multiplexing trade-off for a NC system, in which multiple access to the channel is allowed to be non-orthogonal. On the other hand, our model with orthogonal access of the nodes to the channel does not require a complex successive interference cancellation technique to be implemented at the destination node.

The major goal in this paper is to introduce \textit{practical NC/decoding methods} for improving the diversity order of a network through cooperation with the overall rate of transmission in mind. The contributions of the current paper can be listed as follows. (1) A generalized wireless NC scenario with nodes possessing both relay and source roles and its diversity analysis. (2) Design of novel network codes based on  close-to-optimal linear block codes. (3) Investigation of greedy codes and maximum code rates for desired diversity levels. (4) Application of the SP algorithm for decoding network codes with relay reliability information. The rest of the paper is organized as follows. In Section \ref{sec:wn_model}, we present the wireless network model, the corresponding detection rule that is individually optimal for each user, and a practical enhancement on the proposed network encoding method. We investigate the details of network codes based on linear block codes with emphasis on greedy NC through some sample networks in Section \ref{sec:linear_block}. We also give some asymptotic results based on the rate-diversity order trade-off curves for the proposed NC method and a repetitive method that represents the no NC case. In Section \ref{sec:spnd}, we introduce the SP network decoder that has linear complexity order and yields performance figures very close to that of the optimal decoder. Section \ref{sec:num_results} includes the numerical results for the mentioned network encoding and decoding methods and Section \ref{sec:conc} concludes the paper.

\section{Wireless Network Model}
\label{sec:wn_model}
\subsection{The Network-Coded System and The Corresponding Separation Vector}
\label{sec:sample_nw}
In this work, we analyze a wireless network in which unicast transmission of data symbols, each belonging to a different source, is to be carried out utilizing NC at the intermediate nodes. Under the general operation scheme, every node may act both as a member (source or destination) of a unicast communication pair and as an intermediator (relay) node for other unicast pairs. 

In order to ease the explanation of system model and the roles of nodes in the network, here we start a with a simple network-coded operation depicted in Fig.~\ref{fig:network3}, which makes use of binary NC through usual binary addition operator $\oplus$. The network of interest consists of $k = 3$ source nodes and a dummy node $0$ that represents a hypothetical detector of source packets $\mathbf{u_1}$, $\mathbf{u_2}$, and $\mathbf{u_3}$ at the corresponding destination nodes. The transmission of the these $3$ data packets is allowed to be completed within $n = 4$ orthogonal time slots, which form a round of NC communication with a data rate of $r=\frac{k}{n}=\frac{3}{4} \; packets/transmission \; slot$. The channel is assumed to be shared by a time division multiple access technique for the sake of simplicity in model description and due to causality requirements forcing the intermediate nodes to listen to a symbol before combining it through NC. 

As seen in Fig.~\ref{fig:network3}, the first time slot is reserved for node $1$ to transmit its own data packet $\mathbf{u_1}$ and this transmission is overheard by source nodes $2$ and $3$ in addition to the destination node $0$. We assume that the links between different pairs of nodes are independently Rayleigh faded. The channel corresponding to a link is quasi-static, i.e., constant over a packet and independently faded for different packets. We further assume that there is no feedback of channel state information (CSI) within the system in order to simplify the implementation and that each receiving node, including node $0$, has the perfect knowledge of only the incoming links through measurements of the respective channels. Hence following the transmission of the packet $\mathbf{u_1}$, both node $2$ and node $3$ use respective CSI to obtain their own detection results on the symbols transmitted. Due to the block fading assumption, we consider a single data symbol $u_1$ and its detection/decoding event representing all symbols in the packet. Therefore, corresponding to the detections of $u_1$, each node has also a reliability information based on the probability of error in the detection, which is in fact only a function of its own channel measurement result. In order to counteract the effects of error propagation, this reliability information is passed to the destination node by an intermediate node (node $2$ or $3$), whenever it combines the detected symbol shown by $\hat{u}_1$ with its own and other nodes' symbols. 

In the second time slot, node $2$ transmits its own symbol and this transmission is observed by all other nodes as well. In the following slot, a NC operation is carried out by node $3$, which simply combines its own symbol and its detection result for the first time slot $\hat{u}_1$. In order to inform the destination node $0$, node $3$ has to append the error probability for the network coded symbol $\hat{u}_1 \oplus u_3$ to the packet it formed. In the last slot, once again node $2$ uses the channel to transmit the network encoded data $\hat{u}_1 \oplus u_2$ with its own estimate of $u_1$ and appends the corresponding reliability information to the transmitted packet. Hence the destination node knows only the reliability information for the last two transmissions which incorporate NC, but not the CSI between all intermediate nodes within the system. For the proposed scenario, the overhead of appending reliability information to the network-coded packets on the spectral efficiency is small for large packet lengths. Therefore, the effect of sharing the the reliability information on spectral efficiency is expected to be small.

Up to this point, the sample NC system is detailed in words. From the perspective of destination node $0$, the same system can be described formally using a generator matrix $\mathbf{G}$ (called the transfer matrix in \cite{Xiao10}) and a scheduling vector $\mathbf{v}$. The columns of $\mathbf{G}$, $\mathbf{g}_j's$, represent the combining operations at the intermediate nodes and the entries of $\mathbf{v}$ give the scheduling of the nodes accessing the channel: 
\begin{align}
\label{eqn:gen}
\mathbf{G}=\left[ \begin{array}{cccc}
1 & 0 & 1 & 1 \\
0 & 1 & 0 & 1 \\
0 & 0 & 1 & 0 \end{array} \right], \mathbf{v} = \left[1\ 2 \ 3\ 2\right].
\end{align}

The combined data vector $\mathbf{u}=\left[ u_1\; u_2\; u_3 \right]$ can then be used to form the distributed codeword: $\mathbf{c}=\left[ c_1\; c_2\; c_3\; c_4\right]=\mathbf{u}\mathbf{G}$. The choices $\mathbf{u}$, $\mathbf{G}$, $k$, and $n$ for the parameters defining the operation of network are not arbitrary. They are used intentionally to point out the analogy to regular linear block codes. However, reliable detection of all data symbols, i.e., whole block $\mathbf{u}$, originating from a single error-free source is of interest for a regular decoder; whereas node $0$ may desire to reliably detect, as an example, only $u_1$ under cooperative encoding. Hence we need to identify a parameter that describes the performance for detection of a single symbol $u_1$ as opposed to the codeword $\mathbf{u}$ for our model.

One can show that the minimum distance for $\mathbf{G}$ is $1$. However, we will see that an error event requires at least $2$ bit errors for decoding of $u_1$ at node $0$. Let all the data bits be equal to $0$ without loss of generality, i.e., $\mathbf{u}=\left[ 0\;0\;0\right]$. Hence the transmitted codeword is expected to be $\mathbf{c}=\left[ 0\;0\;0\;0\right]$ for the case of no intermediate node errors. The error event for $u_1$ corresponds to its decoding as $1$. This erroneous decoding can occur for sequence decoding $\hat{\hat{\mathbf{u}}} \in \{[100],[101],[110],[111]\}$, where $\hat{\hat{\mathbf{u}}}$ denotes the decoding result at node $0$. The incorrect codewords $\hat{\hat{\mathbf{c}}}$ corresponding to these decoded vectors are $[1011],[1001],[1110],[1100]$, respectively. When these codewords are compared to the codeword $[0000]$, it is clear that at least $2$ bit errors are needed to cause an error event. Hence the minimum distance for $u_1$ in this setting is said to be $2$. The erroneous decoding for other bits can be investigated in a similar fashion. Focusing on $u_3$ and hypothesizing ${\mathbf{u}}=\left[ 0\;0\;0\right]$, $u_3$ is incorrectly decoded when $\hat{\hat{\mathbf{u}}} \in \{[001],[011],[101],[111]\}$. The corresponding codewords are $[0010],[0111],[1001],[1100]$. Therefore, a single bit error can cause erroneous decoding of $u_3$ yielding a minimum distance of $1$. As seen in the example, the error performance varies from symbol to symbol. Next we generalize this claim to cover arbitrary generator matrices and verify it through simulations in Section \ref{sec:num_results}.
%

Now we consider a subset of nodes in which there are $k$ nodes transmitting data to a single destination node $0$. Let the symbol transmitted by node $i$ be denoted by $u_i$, for $i \in \{ 1, \dots, k\}$, and $u_i$ be an element from the Galois field of size $q$, GF($q$). We assume $u_i$ to be statistically independent and 
define $\mathbf{u}=[u_1 \ u_2 \ \ldots \ u_k]$ as the combined data vector. In time slot $j\in \left\lbrace 1,\ldots ,n\right\rbrace$, a transmitting node $v_j\in \left\lbrace 1,\ldots ,k\right\rbrace$ forms a linear combination of its own and other nodes' data. If $v_j$ detects all data to be encoded correctly, it simply forms $c_j = \mathbf{u}\mathbf{g}_j$, where $\mathbf{g}_j$ is a $k \times 1$ network encoding vector whose entries are elements of GF($q$). Let $\hat{u}_i$ denote the estimate of the symbol of node $i$ at node $v_j$. Using these estimates, node $v_j$ forms the noisy network coded symbol $\hat{c}_j= \mathbf{\hat{u}}\mathbf{g}_j$ that is also an element of GF($q$). Then $v_j$ modulates and transmits this symbol to receiver node $0$ as: 
\begin{equation}
\label{eqn:coded}
s_j=\mu(\hat{c}_j),
\end{equation}
where $\mu(.)$ shows the mapping of a symbol to a constellation point. Although symbols may come from any alphabet and non-binary constellations may be used, we will focus hereafter on GF($2$) and binary phase-shift keying (BPSK) with $s_j = 1-2c_j$. Our assumption is that vector $\mathbf{g_j}$, source address $v_j$ and probability of error $p_{e_j}$ for the transmitted symbol are appended to the corresponding packet. We consider transmissions with no channel coding and deal with single network coded data symbol $c_j$ which represents all symbols within a packet transmitted by $v_j$. At the end of a round of transmissions, if no errors occur at the intermediate nodes, the overall vector of $n$ symbols coded cooperatively in the network is
\begin{equation}
\label{eqn:uG}
\mathbf{c}= \left[c_1\ c_2\ \hdots \ c_n \right] =\mathbf{u} \left[ \mathbf{g}_1 \ \mathbf{g}_2 \ \hdots \ \mathbf{g}_n \right]=\mathbf{u} \mathbf{G}. 
\end{equation}
The generator matrix characterizes the network code together with the vector of transmitting nodes
\begin{equation}
\label{eqn:scheduling}
\mathbf{v}= \left[ v_1\ v_2\ \ldots \ v_n\right].
\end{equation}

Equations (\ref{eqn:uG}) and (\ref{eqn:scheduling}) generalize the definition of the example NC in (\ref{eqn:gen}). Next, we present Algorithm~\ref{alg:mindist}, which generalizes the method for finding the minimum distance for $u_i$. 
\begin{algorithm}
\caption{Algorithm for finding the minimum distance corresponding to symbol $u_i$ for a $(n,k,d)$ code with given generator matrix $\mathbf{G}$}
\label{alg:mindist}
\begin{algorithmic}
\STATE $minimumdistance \gets n$
\STATE $indexvector \gets \left[ 1 \ 2 \ \cdots \ k \right] \setminus i$
\FOR{$j = 1$ to $2^{k-1}$} 
	\STATE $errorpattern \gets dec2GF2(j-1)$ 
	\STATE $errdatavector[i] \gets 1$
	\STATE $errdatavector[indexvector] \gets errorpattern$
	\STATE $errcodevector \gets errdatavector \ast \mathbf{G}$
	\STATE $errcodedistance \gets numberofnonzero(errcodevector)$
	\STATE $minimumdistance \gets min(minimumdistance, errcodedistance)$
\ENDFOR
\end{algorithmic}
\end{algorithm}
In Algorithm \ref{alg:mindist}, the function $dec2GF2(.)$ returns a binary pattern corresponding to the input decimal number and the function $numberofnonzero(.)$ returns the number of non-zero entries in the input vector. It is assumed that the data vector $\mathbf{u}$ consists of all $0$s, relying on the linearity of the network code. The algorithm first creates all possible erroneous data vectors $\mathbf{\hat{\hat{u}}}$ that have $1$ in the $i$th position so that all possible codewords leading to erroneous decoding of $u_i$ are generated by $\hat{\hat{\mathbf{c}}}=\mathbf{\hat{\hat{u}}}\mathbf{G}$. Afterwards, we search within these codewords to find the one with the minimum distance to the transmitted codeword of all $0$s. This minimum value gives us the minimum distance for $u_i$. The set of minimum distances corresponding to all $u_i$'s (named the separation vector \cite{Gils83}) is utilized in identifying the performance metrics for NC in Section~\ref{sec:linear_block_diversity}.

\subsection{Optimal Network Decoding Using Intermediate Node Reliability Information}
\label{sec:wn_model_optimal}
The intermediate nodes are assumed to use the DetF technique (hard decision with no decoding operation) due to its simplicity. In a wireless network, an intermediate node $v_j$ has a noisy detection result $\hat{\mathbf{u}}$ of $\mathbf{u}$. Let us express the resulting noisy network coded symbol as
\begin{equation}
\label{eqn:noisy2}
\hat{c}_j=c_j\oplus e_j,
\end{equation}
where $e_j$ denotes this propagated error. We observe that a possible error in $\hat{\mathbf{u}}$ propagates to $\hat{c}_j$ after the NC operation dictated by $\mathbf{g}_j$ is realized. We assume that node $v_j$ knows the probability mass function (PMF) of $e_j$, $p(e_j)$, which we name as the intermediate node reliability information. This assumption is not unrealistic as it can be determined by the estimation of the channel gains of the links connected to $v_j$, along with the reliability information forwarded to $v_j$. The received signal by node $0$ at time slot $j$ is then $y_j=h_js_j+w_j$, where $h_j$ is the channel gain coefficient resulting from fading during the $j$th slot and $w_j$ is the noise term for the link between $v_j$ and node $0$. The gain coefficient is circularly symmetric complex Gaussian (CSCG), zero-mean with variance $E_s$, i.e., it has distribution $\mathbb{CN}(0,E_s)$. The noise term is CSCG with $\mathbb{CN}(0,N_0)$. The usual independence relations between related variables representing fading and noise terms exist. The overall observation vector of length $n$ at node $0$ is
\begin{align}
\label{eqn:obs}
\mathbf{y}=\mathbf{H} \mathbf{s}+\mathbf{w},
\end{align}
where ${\mathbf{y}}=[y_1 \ \ldots \ y_n]^T, {\mathbf{s}}=[s_1 \ \ldots \ s_n]^T=\mu (\hat{\mathbf{c}}^T), {\mathbf{w}}=[w_1 \ \ldots \ w_n]^T$ and $\mathbf{H}$ is a diagonal matrix whose elements are independent channel gains $h_1,h_2,\ldots,h_n$ for the links connected to node $0$. It is assumed that $\mathbf{H}$ is perfectly known at node $0$. Combining the coded symbols in a network code vector, we obtain
\begin{equation}
\label{eqn:codevect}
\mathbf{\hat{c}} = \mathbf{c} \oplus \mathbf{e} = \mathbf{uG} \oplus \mathbf{e},
\end{equation}
where $\mathbf{e} = \left[ e_1 \ \ldots \ e_n\right]$ is the error vector denoting the first hop errors. We assume that $\mathbf{e}$ is independent of $\mathbf{c}$ although dependence can be incorporated in the SP decoder developed in Section \ref{sec:spnd}. This independence assumption is valid directly for BPSK modulation, whereas in a general modulation scheme the Euclidean distances between various constellation point pairs differ and an error term $e_j$ depends on the symbol being transmitted. As a result, using (\ref{eqn:noisy2}), (\ref{eqn:obs}), and  (\ref{eqn:codevect}), the observation vector at node $0$ is
\begin{equation}
\label{eqn:obs2}
\mathbf{y} = \mathbf{H} \ \mu(\mathbf{uG} \oplus \mathbf{e})^T +\mathbf{w}.
\end{equation}
Thus node $0$ has access to the likelihood $p(\mathbf{y} \vert \mathbf{u,e})$ and $p(\mathbf{e})=\prod_{j=1}^{n}p(e_j)$, assuming the errors are independent. As shown in \cite{Laneman06}, in order to avoid the propagation of errors occurring at intermediate nodes, node $0$ has to utilize the reliability information $p(\mathbf{e})$. Then, the a posteriori probability of the source bit of interest, say $u_1$, can be calculated by using the Bayes' rule:
\begin{align}
p(u_1 \vert \mathbf{y}) =\alpha \sum_{u_2,\ldots ,u_k}\sum_{e_1,\ldots ,e_n} \; p(\mathbf{y} \vert \mathbf{u,e})\prod_{j=1}^{n}p(e_j),
\end{align}
where $\alpha$ is a normalizing constant that does not depend on $u_1$. The MAP estimate of $u_1$ at node $0$ is denoted by $\hat{\hat{u}}_1$ and obtained as 
\begin{align}
\label{eqn:opt_det}
\hat{\hat{u}}_1 = \argmax_{u_1}p(u_1\vert \mathbf{y})= \argmax_{u_1}\sum_{u_2,\ldots ,u_k}\sum_{e_1,\ldots ,e_n} p(\mathbf{y} \vert \mathbf{u,e})\prod_{j=1}^{n}p(e_j),
\end{align}
which is the individually optimum detector for $u_1$. As a result, for the optimal detection of $u_1$, node $0$ requires the intermediate node reliability information vector: $\mathbf{p_{e}} = [p_{e_{1}}\ \ldots \ p_{e_{n}}]$, where  $p_{e_{j}}=P(e_j = 1)$  depends on the PMF of $e_j$. We observe the performance of this detection rule in Section \ref{sec:num_results_sub1}.

The problem related to the MAP-based detection rule of (\ref{eqn:opt_det}) is the number of required operations, which grows exponentially both in the number of nodes $k$ and the number of possible error events $n$. This is addressed in Section \ref{sec:spnd}, where we suggest a practical network decoding technique.

\subsection{Selective Network Coding}
\label{sec:snc}
The NC described in Section II-A is a static method in the sense that the generator matrix $\mathbf{G}$ is fixed. In static NC, node $v_j$ always combines (network encodes) the symbols of a pre-determined set of users, even when it knows that the reliability for one of those users is low. When a symbol estimate with low reliability is combined with a symbol with high reliability, the reliability of the resulting network coded symbol is low. Thus, it is intuitive to expect some gains in performance by forcing the intermediate nodes not to combine the symbols that have very low instantaneous reliability. In \cite{Herhold04} and \cite{Onat08}, various forms of channel state information are used to determine thresholds for relaying decisions. In \cite{Xiao10} and \cite{Hanzo11}, for relays assuming DF operation, successful decoding of channel code for a source is the required condition for combining its data in NC. Here, we propose a method called Selective Network Coding (SNC) that imposes a threshold on the reliability of the candidate symbols to be encoded at intermediate nodes that adapt DetF. In this way, any symbol that is sufficiently reliable is included in network encoding and the resulting encoding vector $\mathbf{g}_j$ is appended to the transmitted packet so that node $0$ still has the instantaneous generator matrix $\mathbf{G}$ at the end of $n$ transmissions.

Let us demonstrate the operation under SNC on the sample network given in Fig.~\ref{fig:network3}. For the first two time slots SNC is equivalent to NC since no combining of other nodes' symbols is the case. However, in the third slot, node $3$ checks the reliability of the detection for $u_1$ carried out following the first slot. Let us say it has observed an instantaneous SNR value of $\gamma _{1\to 3}$, which yields a probability of error equal to $p_{e_3} = Q(\sqrt{2\gamma _{1\to 3}})$ in detection of $u_1$ for BPSK modulation. Here, the Q-function is defined as $Q(x)=\frac{1}{\sqrt{2\pi}} \int_x^\infty exp(-\frac{z^2}{2})\text{d}z$ and the random variable $\gamma _{1\to 3}$ is exponentially distributed with mean value equal to average SNR $\bar{\gamma}$ for a Rayleigh fading channel. The instantaneous error probability $p_{e_{3}}$ is a measure of reliability for $\hat{u}_1$. This instantaneous $p_{e_3}$ value is averaged over $\gamma _{1\to 3}$ to set the threshold:
\begin{align}
p_{th_3}=&\text{E}_{\gamma _{1\to 3}}\left\lbrace p_{e_3} \right\rbrace 
		= \int_0^\infty Q(\sqrt{2\gamma _{1\to 3}})\frac{1}{\bar{\gamma}}\exp\left( -\frac{\gamma _{1\to 3}}{\bar{\gamma}} \right)\text{d}\gamma _{1\to 3}\nonumber\\
		=&\frac{1}{2}\left( 1 - \sqrt{\frac{\bar{\gamma}}{1 + \bar{\gamma}}}\right),
\end{align}
where $\text{E}_{\gamma _{1\to 3}}\left\lbrace.\right\rbrace$ denotes the expectation operator over $\gamma _{1\to 3}$. Therefore, node $3$ uses the threshold value $p_{th_3}$ to check whether the detection at that instant is reliable. If $p_{e_3}<p_{th_3}$, the detection is decided to be reliable enough and the combination $\hat{u}_1 \oplus u_3$ is formed just in the way declared by the generator matrix $\mathbf{G}$. Otherwise, node $3$ modulates and transmits only its own symbol $u_3$ and appends this information to the corresponding packet. Similarly, in the last slot, node $2$ checks the reliability of its own detection of $u_1$ and forms either $\hat{u}_1 \oplus u_2$ or simply transmits $u_2$. Here, the reliability of $\hat{u}_1 \oplus u_2$ is equal to the reliability of $\hat{u}_1$. In general, there may be more than one symbol that an intermediate node should detect and combine according to $\mathbf{G}$. In such cases the combined instantaneous reliability of an network encoded symbol at the time slot $j$ can be easily obtained by
\begin{align}
\label{eqn:combined_err}
p_{e_j} = \frac{1-\prod_{i\in A_j}\left(1-2P_j(\hat{u}_i\neq u_i)\right)}{2},
\end{align}
where $A_j$ denotes the set of sources for which $v_j$ should carry out the network coding, i.e., $A_j$ is the set of indices corresponding to the non-zero elements of the $j$th column of $\mathbf{G}$, $\mathbf{g}_j$. The term $P_j(\hat{u}_i\neq u_i)$ in (\ref{eqn:combined_err}) is used to show the probability of error for detection of $u_i$ by node $v_j$.

Clearly, SNC inherently includes usage of adaptive generator matrices. The utilized generator matrix may assume in average a form dictated by some predetermined (and optimal if possible) linear block code structure like the ones that are to be discussed in Section \ref{sec:greedy}. Note that we do not claim the optimality of the proposed threshold. However, the bit error rate (BER) performance improvements are observed in Section~\ref{sec:perf_snc}. 

\section{Linear Block Codes Utilized as Network Codes}
\label{sec:linear_block}
\subsection{Using Separation Vector as a Performance Metric}
\label{sec:linear_block_diversity}
Our goal is now to explore the error performance metrics for network coding/decoding described in Section \ref{sec:wn_model}. Our basic figure of merit will be the diversity order corresponding to the source bit $u_i$, which is an asymptotic term defined for SNR tending to infinity:
\begin{align}
d_i = -\lim_{SNR\to\infty}\frac{\log P_{\hat{\hat{u}}_i\neq u_i}(SNR)}{\log SNR}
\end{align}
giving information on the slope of decrease in logarithm of BER for $u_i$, i.e., $P_{\hat{\hat{u}}_i\neq u_i}$ for high SNR values. For conventional block coding, the average error performance over all data symbols is of interest. Therefore, for a linear block code whose coded symbols are transmitted over independent channels, the metric utilized for comparison is the minimum distance, which is equal to the diversity order \cite{Proakis}. On the other hand, there is a vector of distinct minimum distances, i.e, separation vector, for data symbols, whenever we are interested in performance of individual symbols that originate from different source nodes. According to the results presented for suboptimal decoders in \cite{Renzo11c, Renzo11d}, the diversity orders for symbols in some sample NC systems are still equal to the minimum distances in the corresponding separation vector in spite of the inherent error propagation problem. Using the soft decoding that we propose in (\ref{eqn:opt_det}) and also authors analyze in \cite{Renzo11b}, one should expect better performance and consequently diversity orders being equal to the minimum distances. A similar result is also shown for a simpler cooperative network with possible relay errors and the use of \textit{equivalent channel} defined as the combination of the source-to-relay and the relay-to-destination channels \cite{Wang07}. In \cite{Wang07}, even a suboptimal detection rule utilizing this equivalent channel approach is shown to attain the achievable diversity order. As a result, supported with intermediate node reliability information, the optimal rule of (\ref{eqn:opt_det}) given in Section~\ref{sec:wn_model_optimal} also satisfies the diversity orders dictated by the separation vector whose entries are obtained according to Algorithm~\ref{alg:mindist}. It should be also noted that since diversity order is an asymptotic quantity, the exact form of $\mathbf{v}$ is irrelevant to the procedure used for obtaining a diversity order value. On the other hand, it is wiser that each column $\mathbf{g_j}$ of $\mathbf{G}$ is used as the encoding function for a $v_j$ such that the $j$th entry is non-zero, $\mathbf{g_j}(v_j) \neq 0$. Otherwise possibly an extra relaying error is also included in the encoded data symbol. Therefore, $\mathbf{v}$ clearly affects the coding gain corresponding to the BER versus SNR curve of $u_i$. 
 
\subsection{An Example of Close-to-Optimal Linear Block Codes: Greedy Codes}
\label{sec:greedy}
In this study, we make use of some well-known linear block codes while constructing network codes that are to be used for the analysis of data rate and diversity orders for distinct symbols in Section \ref{sec:rate_div} and simulation of BER in Section \ref{sec:num_results}. However, the cooperative network coded operation described in this work and the resulting performance figures for a unicast pair are more general and applicable to any linear block code like the maximum distance separable (MDS) codes detailed in the context of NC in \cite{Xiao10}. 

In comparison with the network coded operation, we consider a case with no distributed coding (no network coding) among the nodes. For this no network coding scenario, we should also consider that our system model does not allow feedback of CSI within the network and that the average SNR values between all nodes are equal. If one intends to achieve higher diversity orders, two resources are available in such a scenario: (i) the temporal diversity resources over the faded blocks, (ii) the spatial diversity resources over the intermediate nodes. Here, it is seen that the source nodes must simply repeat their data instead of choosing a relay to convey their data which may possibly inject errors leading to worse performance than repetition. In conclusion, we call this method as the repetition coding scheme which is in fact a degenerate NC scheme with no cooperation hence with reduced spatial diversity resources. Following $n$ transmissions, node $0$ combines the data received for each source symbol optimally to generate the detection results. 

On the other hand, with NC, we take the family of block codes known as greedy codes as an example. These ($n$, $k$, $d$) codes are selected with the following parameters: blocklength (number of transmission slots) $n$, dimension (number of unicast pairs) $k$, and minimum distance (minimum diversity order) $d$. Greedy codes are known to satisfy or be very close to the optimal dimensions for all blocklength-minimum distance pairs \cite{Pless93} and can be generalized to non-binary fields \cite{Monroe95} for achieving higher diversity orders with NC as discussed in \cite{Xiao10}. Moreover, they are readily available for all dimensions (number of nodes) and minimum distances unlike some other optimal codes. Hence, even in an ad hoc wireless network with time-varying size, any desired diversity order can be flexibly satisfied by simply broadcasting the new greedy code generator matrix $\mathbf{G}$ to be utilized in subsequent rounds of communication. 

As an example, let us consider a network that consists of $k=3$ nodes transmitting their data symbols over GF($2$). If a round of communication is composed of $n=6$ transmission slots, we deal with codes of type ($6$, $3$, $d$), which have a code rate of $\frac{1}{2}$. Starting with the generator matrix and scheduling vector corresponding to the repetition coding, we have
\begin{align}
\label{eqn:gen2}
\mathbf{G}&=\left[ \begin{array}{cccccc}
1 & 0 & 0 & 1 & 0 & 0 \\
0 & 1 & 0 & 0 & 1 & 0 \\
0 & 0 & 1 & 0 & 0 & 1 \end{array} \right], \mathbf{v}_{\phantom{1}} = \left[1\ 2 \ 3\ 1\ 2 \ 3\right].
\end{align}
It is easily observed that, since each data bit is transmitted twice over independent channels, this method satisfies only a diversity order of $2$ for all bits $u_1$, $u_2$, and $u_3$. In contrast, a diversity order of $3$ for all sources can be achieved using NC, with the same code rate. As an example, the NC that achieves this performance can be obtained using the ($6$, $3$, $3$) greedy code, as follows:
\begin{align}
\label{eqn:gen3}
\mathbf{G}_1&=\left[ \begin{array}{cccccc}
1 & 0 & 0 & 1 & 1 & 0 \\
0 & 1 & 0 & 0 & 1 & 1 \\
0 & 0 & 1 & 1 & 0 & 1 \end{array} \right], \mathbf{v}_1 = \left[1\ 2 \ 3\ 1\ 2 \ 3\right].
\end{align}
Clearly, without NC, the diversity order of $3$ for all sources can only be achieved with rate $\frac{1}{3}$. It should also be noted that greedy codes accommodate each unicast pair with equal diversity order due to the greedy algorithm utilized in their construction. Moreover, contrary to the findings in \cite{Xiao10}, it is easy to obtain any required diversity order for any data bit even by using GF($2$). The limitation is not due to the  number of unicast pairs but due to the number of transmission slots in general. By increasing $n$, one can arrange and improve the diversity orders, if the transmissions to each node are realized over independent channels, which is a natural assumption for many wireless communication scenarios. If we need an increase in data rate, through a trade-off mechanism, we can assign decreased diversity orders to the lower-priority unicast pairs. This may be accomplished by omitting some columns of a greedy code generator matrix in order to decrease number of transmissions. The columns to be excluded can be decided by running Algorithm \ref{alg:mindist} in Section \ref{sec:sample_nw} on candidate punctured generator matrices. As an example, the following punctured ($5$, $3$, $2$) code is obtained by omitting the last column of $\mathbf{G}_1$ and has a data rate $\frac{3}{5}$ that is higher than those of above two codes:
\begin{align}
\label{eqn:gen4}
\mathbf{G}_2=\left[ \begin{array}{ccccc}
1 & 0 & 0 & 1 & 1  \\
0 & 1 & 0 & 0 & 1  \\
0 & 0 & 1 & 1 & 0  \end{array} \right], \quad\ \mathbf{v}_2 = \left[1\ 2 \ 3\ 1\ 2 \right].\quad
\end{align}
This punctured network code satisfies a diversity order of $3$ for $u_1$ and an order of $2$ for both $u_2$ and $u_3$. If $u_1$ is of higher priority, this unequal error protection would be preferable especially when the higher rate of the code is considered. In case of larger diversity order requirement, $d=4$ as an example, we may simply utilize the ($7$, $3$, $4$) greedy code with rate $\frac{3}{7}$.
\begin{align}
\label{eqn:gen5}
\mathbf{G}_3&=\left[ \begin{array}{ccccccc}
1 & 0 & 0 & 1 & 1 & 0 & 1\\
0 & 1 & 0 & 0 & 1 & 1 & 1\\
0 & 0 & 1 & 1 & 0 & 1 & 1\end{array} \right], \mathbf{v}_3 = \left[1\ 2 \ 3\ 1\ 2 \ 3 \ 1\right].
\end{align}
A final problem is the selection of vector $\mathbf{v}$. Our basic assumption is that $\mathbf{v}$ satisfies causality so that no intermediate node $v_j$ tries to transmit another node's symbol before hearing at least one copy of it. This causality problem can be solved trivially by using only systematic generator matrices. For the transmitting nodes corresponding to the non-systematic part of $\mathbf{G}$, as described in Section \ref{sec:linear_block_diversity}, one can select each entry $v_j$ such that $\mathbf{g_j}(v_j) \neq 0$ for each column $\mathbf{g_j}$. For the columns that have more than one non-zero entry, a random selection between candidate $v_j$'s will merely affect the coding gains assigned to these nodes. As a result, one can force the number of transmissions of each node within a round to be equalized as much as possible for similar coding gain improvements of nodes. In the way exemplified in this section, one can choose a network code satisfying desired error protection properties for a determined network size with adequate data rate quite flexibly.

\subsection{Theoretical Gains in Rate and Diversity for NC}
\label{sec:rate_div}
\begin{figure}[!t]
\centering
\includegraphics[width=\columnwidth]{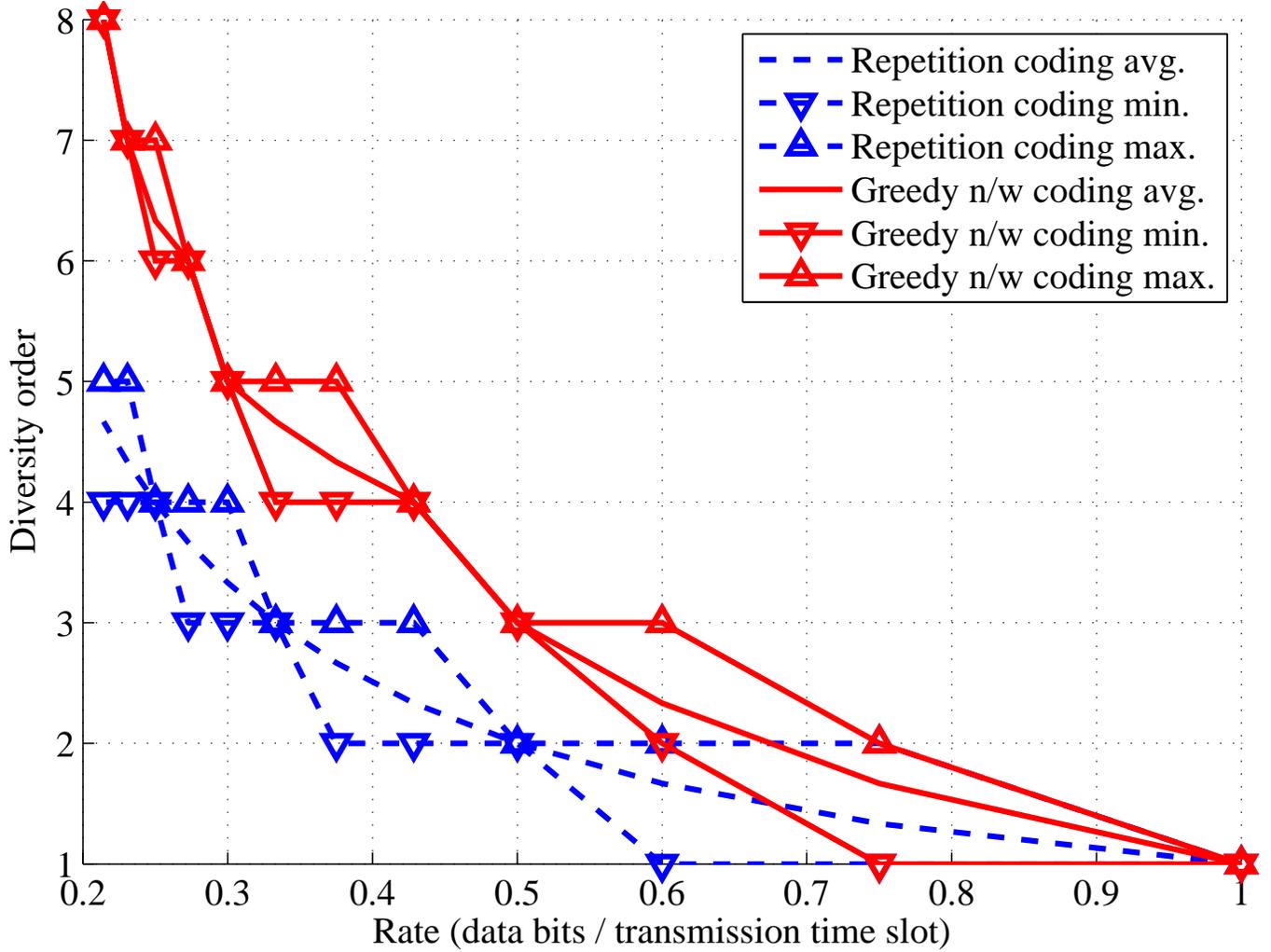}
\caption{Network diversity orders for greedy NC and repetition coding for given data rates.}
\label{fig:ber_rate_div}
\end{figure}

In this section, we investigate the rate and diversity (asymptotic) gains of NC through use of the family of greedy network codes detailed in Section \ref{sec:greedy}, although the results are still valid for any other family of optimal or close-to-optimal codes. The availability of a greedy code for a given $(k,d)$ pair is checked using \cite{BobJenkins}. Fig.~\ref{fig:ber_rate_div} shows the diversity gains attainable using greedy NC (with punctured codes in case no corresponding greedy code exists) with respect to the repetition coding scenario. The rate-diversity trade-off curves of both cases are plotted for a network of $k=3$ nodes with increasing number of transmissions and hence decreasing rate. We are interested in three types of network diversity orders; average, minimum and maximum, since the orders corresponding to each one of the three nodes may be unequal in general. The curves with no markers represent the (average) network diversity orders for both scenarios, which is defined as the arithmetic mean of orders for three nodes. For a rate of $0.43\ \text{bits/transmission}$, with greedy code $(7,3,4)$, the network diversity order for NC is $4$. The minimum, maximum, and average diversity orders are equal for this case. In contrast, the repetition scheme results in an average order of nearly $2.33$ with the worst node observing a minimum order of $2$ and the best node a maximum order of $3$, which would mean a high SNR loss asymptotically for all three nodes in the network. 

\begin{figure}[!t]
\centering
\includegraphics[width=\columnwidth]{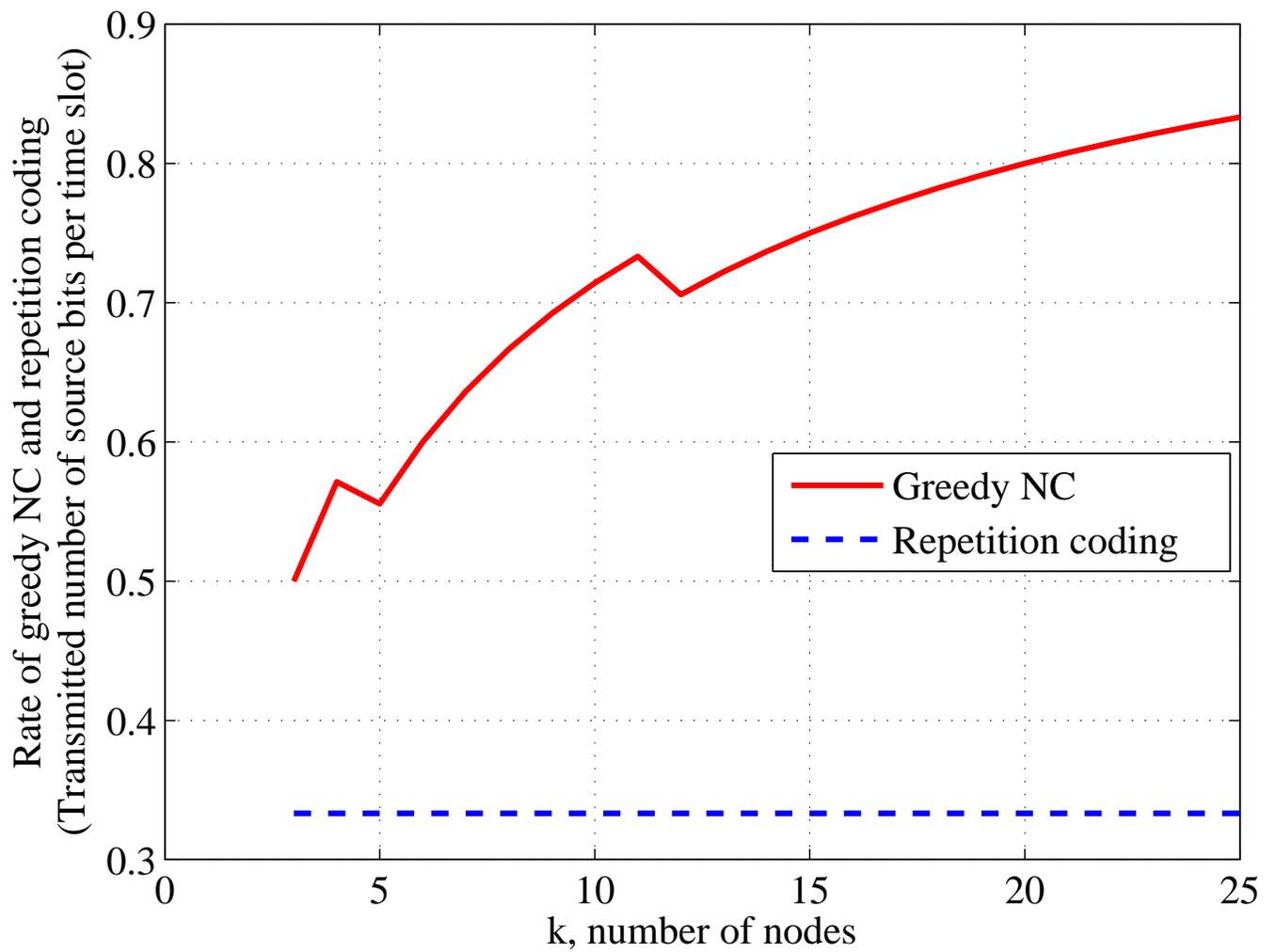}
\caption{Increasing rate advantage of greedy NC for increasing network size.}
\label{fig:ber_rate_adv}
\end{figure}

In Fig.~\ref{fig:ber_rate_adv}, we now fix the desired network diversity order to $d=3$ and observe the rate advantage of the NC for increasing network size. Note that for all cases diversity orders for $k$ users are equal to $3$. For a network of $k=25$ nodes, the rate with NC is $\frac{25}{30}$ (with greedy code $(30,25,3)$) and the rate of the repetition scheme is $\frac{15}{45}$ (always equal to $\frac{1}{3}$ for a diversity order of $3$). The rate advantage ratio is then $2.5$. In the asymptotic case, as $k\rightarrow \infty$ and hence as $n\rightarrow \infty$, NC using optimal codes in construction will have a rate advantage converging to $3$ since the rate for network coded case can be shown to tend to $1$ using the Gilbert-Varshamov bound \cite{Wicker} for arbitrarily large $n$. In general, the rate advantage of NC over the repetition scenario becomes simply $d$, the desired network diversity order. As a result, increasing the network size improves the network coded system's efficiency in comparison to the repetition coding. 

\section{Sum-Product Network Decoder}
\label{sec:spnd}
It is clear that the complexity of the optimal rule for decoding of any unicast transmission symbol $u_i$ grows exponentially, since the number of additions and multiplications in (\ref{eqn:opt_det}) increase exponentially in the number of users $k$ and transmissions $n$. Therefore, this rule becomes quickly inapplicable even for moderate-size networks. Recently the SP iterative decoding, which is often utilized for decoding of low-density parity-check (LDPC) codes, is suggested for decoding general linear block codes as well \cite{Moon04}.

\begin{figure}[!t]
\centering
\includegraphics[width=\columnwidth]{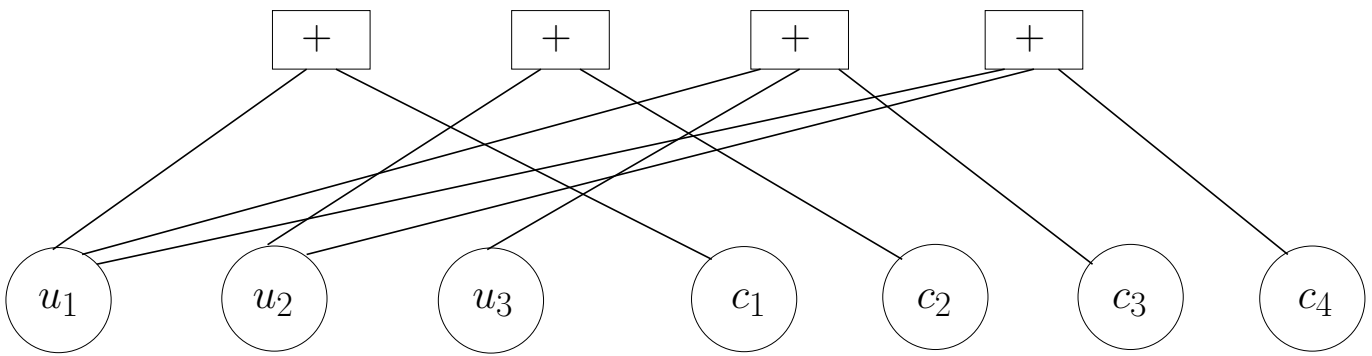}
\caption{Tanner graph for network coded system of (\ref{eqn:gen})}
\label{fig:tanner}
\end{figure}

Here, under the Rayleigh fading scenario detailed in Section \ref{sec:wn_model_optimal}, we make use of SP decoding and compare its performance with that of the optimal one. The aim of the decoding operation is to produce a posteriori probabilities (APPs) for source symbols $u_1,\ldots, u_k$. To that end, we form a combined codeword $\left[ u_1 \ldots u_k \; c_1 \ldots c_n\right]$ and consider the parity check matrix for this codeword, which describes the underlying linear block code structure of the network code. On the Tanner graph, we add a variable node for each source symbol $u_i$, $i=1,\ldots, k$ and each coded symbol $c_j$, $j=1,\ldots, n$.  Afterwards, we add the check nodes which reflect the connections between the source and the coded symbols in the way described by the parity check matrix. For the NC system given in (\ref{eqn:gen}), we refer to the graph presented in Fig.~\ref{fig:tanner} for SP decoding at node $0$. The parity check matrix for this system becomes:
\begin{align}
& \begin{array}{ccccccc}
\phantom{\left[ \right.} u_1 & u_2 & u_3 & c_1 & c_2 & c_3 & c_4 \phantom{\left. \right] }
\end{array}\nonumber\\
\begin{array}{c}
\text{Parity Check 1} \to\\
\text{Parity Check 2} \to\\
\text{Parity Check 3} \to\\
\text{Parity Check 4} \to\\
\end{array}
& \left[ \begin{array}{ccccccc}   
 1\,\, & 0\,\, & 0\,\, & 1\,\, & 0\,\, & 0\,\, & 0\,\,\\
 0 & 1 & 0 & 0 & 1 & 0 & 0\\
 1 & 0 & 1 & 0 & 0 & 1 & 0\\
 1 & 1 & 0 & 0 & 0 & 0 & 1\end{array} \right] = \left[ \mathbf{G}^T \; \vdots \; \mathbf{I}_n \right],
\end{align}
where $\mathbf{G}^T$ denotes the transpose of $\mathbf{G}$ and $\mathbf{I}_n$ is the $n\times n$ identity matrix. 
For a regular LDPC decoder, all of the variable nodes are observed through the channel and corresponding to each channel observation an LLR is computed. For our case, the variable nodes $u_1$, $u_2$, and $u_3$ are not observed so the corresponding LLRs are set to $0$. The channel LLRs for the remaining nodes ($c_1$, $c_2$, $c_3$, and $c_4$) cannot be calculated as in a regular LDPC decoder either, due to the intermediate node error events. Taking these errors into account by using (\ref{eqn:noisy2}) and (\ref{eqn:obs2}), the channel LLR of $c_j$ is: 
\begin{align}
\text{LLR} (c_j) &= \ln \frac{p(y_j\vert c_j = 0)}{p(y_j\vert c_j = 1)}\nonumber\\
&= \ln \frac{(1-p_{e_j})p(y_j \vert \hat{c}_j=0) + p_{e_j}p(y_j \vert \hat{c}_j=1)} {(1-p_{e_j})p(y_j \vert \hat{c}_j=1) + p_{e_j}p(y_j \vert \hat{c}_j=0)}\nonumber\\
&= \ln \frac{\exp (\text{LLR}(e_j)) \exp (\text{LLR}(\hat{c}_j)) + 1}{\exp (\text{LLR}(e_j)) + \exp (\text{LLR}(\hat{c}_j))},
\end{align}
where
\begin{equation}
\text{LLR}(e_j) \triangleq \ln \frac{1-p_{e_j}}{p_{e_j}} \; \text{and LLR}(\hat{c}_j) \triangleq \ln \frac{p(y_j \vert \hat{c}_j=0)}{p(y_j \vert \hat{c}_j=1)} = \frac{4\mathfrak{Re} \left\lbrace h_j^* y_j \right\rbrace}{N_0},
\end{equation}
where $h_j^*$ is the conjugated gain of the channel over which the modulated symbol $s_j = \mu (\hat{c}_j)$ is transmitted by node $v_j$ and we use the fact that $w_j$ is Gaussian distributed (see Section \ref{sec:wn_model_optimal}) in obtaining LLR$(\hat{c}_j)$. 
Given the channel LLRs, the SP decoder carries on iterations over the Tanner graph to generate the estimated LLRs for the source bits. If the number of iterations is fixed, the SP decoder utilized is known to have a complexity order of $O(n)$. In contrast, the optimal decoder has a computational load in the order of $O(2^n)$, which makes the SP network decoder a strong alternative for increasing network size and number of transmissions. One may also note that the proposed scheme works directly with GF($q$), $q>2$, and constellations other than BPSK. The use of higher order fields and constellations would tremendously increase the complexity of the optimal algorithm and make it impractical, whereas the SP algorithm would still operate with reasonable complexity. The number of iterations and other operational parameters for the SP decoder are given in Section \ref{sec:perf_sum}, where we show that performance figures close to that of the optimal one are possible for the network codes investigated herein.

\section{Numerical Results}
\label{sec:num_results}
\subsection{Sample Network-I: Simulation Results}
\label{sec:num_results_sub1}
\begin{figure}[!t]
\centering
\includegraphics[width=\columnwidth]{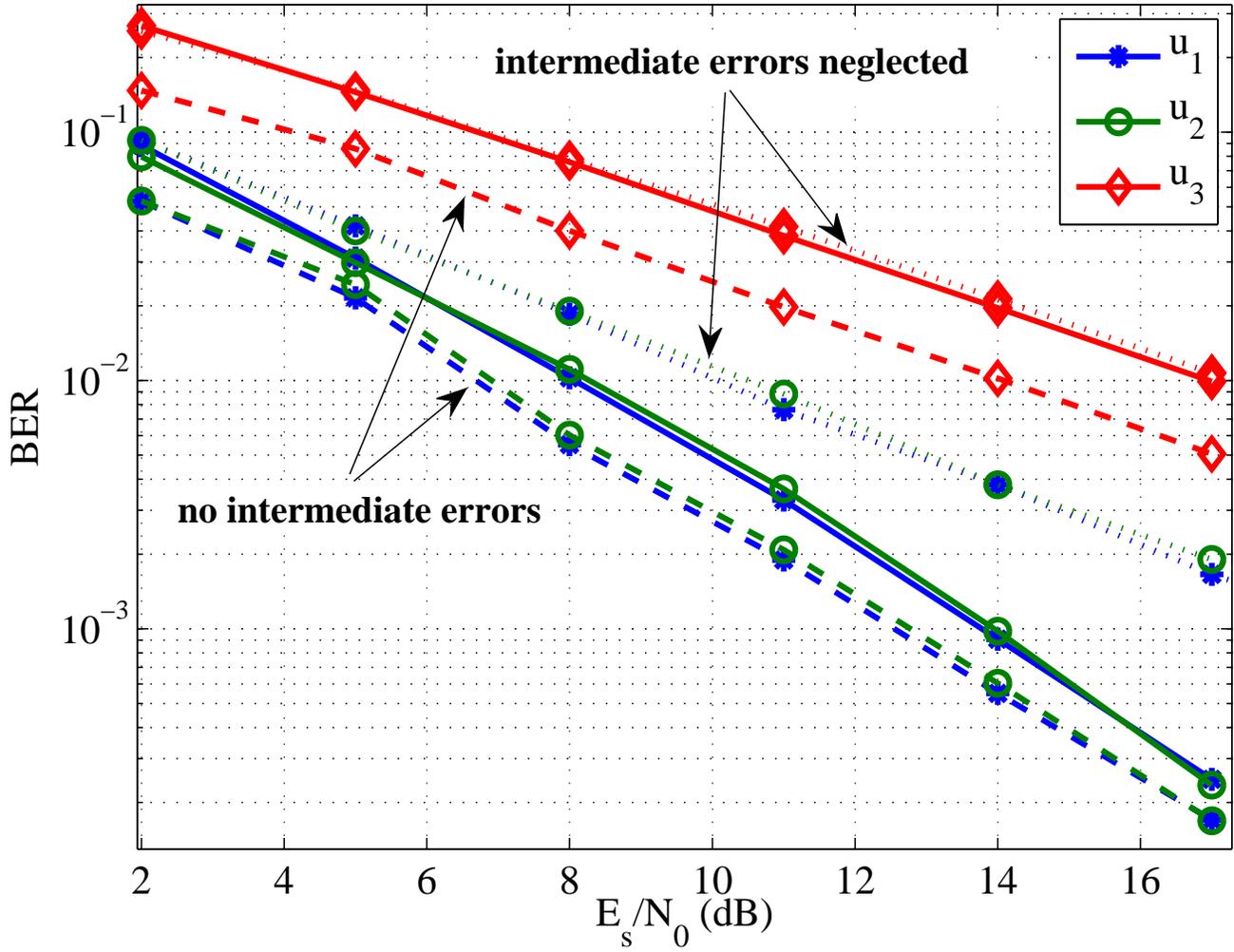}
\caption{BER performance for data bits of different nodes for optimal detection.}
\label{fig:ber_optimum}
\end{figure}

The results in this subsection are based on Sample Network-I of (\ref{eqn:gen}), consisting of only $4$ nodes in order to observe the fundamental issues. For BER results, at least $100$ bit errors for each data bit $u_1$, $u_2$, and $u_3$ are collected through Monte Carlo simulations for each SNR value. In each run, data bits, intermediate node errors and complex channel gains are randomly generated with their corresponding probability distributions. The solid lines in Fig.~\ref{fig:ber_optimum} show the BER values for the optimal detector operating under the realistic scenario of intermediate node errors, whereas the dashed lines depict the performance of the genie-aided no-intermediate-error network with the same optimal detection. Finally, the dotted lines are for the detector that totally neglects possible intermediate errors. 
 
It is observed in Fig.~\ref{fig:ber_optimum} that different diversity orders for bits of different nodes are apparent for optimal detection under intermediate errors. The diversity order for $u_1$ is observed to be $2$ according to the slope of the corresponding BER curve. This is in agreement with the analytical results in Section \ref{sec:sample_nw} where it was shown that an error event corresponds to at least $2$ bit errors for the detection of $u_1$ and $u_2$. It is seen in Fig.~\ref{fig:ber_optimum} that the intermediate node errors cause no loss of diversity for $u_1$ and $u_2$, but an SNR loss of $1.5$ dB. Hence the optimal detection rule of (\ref{eqn:opt_det}) is said to avoid the problem of error propagation in terms of the diversity orders. The loss for $u_3$, whose diversity order is $1$, with respect to the hypothetical no-intermediate-error network is around $2.5$ dB. The performance deteriorates significantly for especially $u_1$ and $u_2$ when intermediate errors are neglected in detection (dotted lines), i.e., $p_{e_3}=p_{e_4}=0$ is assumed. Not only an SNR loss is endured but also the diversity gains for them disappear.
\subsection{Sample Network-II: Simulation Results}
\label{sec:num_results_sub2}
Next, we verify the analytical results concerning the diversity orders for a set of three nodes operating under three different network codes constructed in Section \ref{sec:greedy}. Moreover, the unequal error protection performance of one of these codes is identified together with the rate advantage it provides.

\begin{figure}[!t]
\centering
\includegraphics[width=\columnwidth]{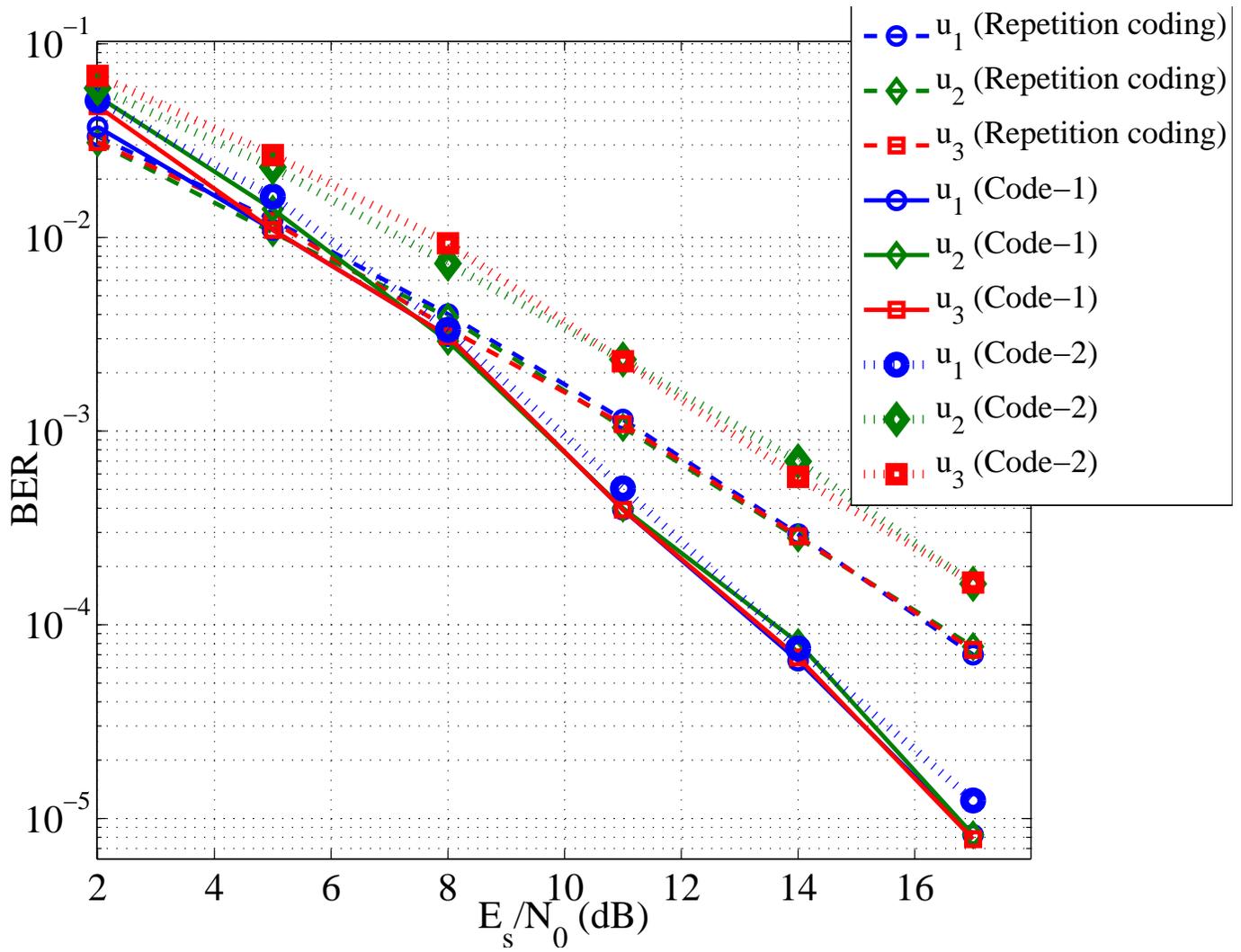}
\caption{BER performance for repetition coding and NC with greedy codes.}
\label{fig:ber_6_3}
\end{figure}

The repetition method is represented by $\mathbf{G}$ and $\mathbf{v}$ in (\ref{eqn:gen2}). To construct Code-1 and Code-2, we make use of the greedy code of (\ref{eqn:gen3}) and the punctured greedy code of (\ref{eqn:gen4}) respectively. Fig.~\ref{fig:ber_6_3} exhibits the BER curves for the repetition scenario with $n=6$ transmissions (dashed lines), for NC scenarios with Code-1 with $n=6$ (solid lines) and Code-2 with $n=5$ (dotted lines). The optimal detector of (\ref{eqn:opt_det}) is utilized for this simulation. Clearly, Code-1 has superior performance with an average network diversity order of $3$. However, the lower rate of Code-1 (and also repetition coding) in comparison to Code-2 should also be noted. For Code-2, on the other hand, bits $u_2$ and $u_3$ observe a diversity order of $2$ while $u_1$ observes an order of $3$. With this unequal protection in mind, the average network diversity order for Code-2 is $\frac{2+2+3}{3}\simeq 2.33$, which is higher than that of the repetition coding with order $2$. In addition to improved diversity, Code-2 has also the advantage of increased overall rate and decreased decoding delay due to usage of $5$ slots instead of $6$. It is preferable especially for a network that puts higher priority on $u_1$.
\subsection{Performance of the Sum-Product Decoding for Network Coded Systems}
\label{sec:perf_sum}
In this section the performance figures for the SP iterative network decoder described in Section \ref{sec:spnd} are presented in comparison with the optimal detection rule of (\ref{eqn:opt_det}), which has an exponential complexity order. The network coded communication system of interest is given in (\ref{eqn:gen3}). The number of iterations for the SP type decoder is limited to $4$ and no early termination is done over parity checks. Here, a minimum of $150$ bit errors are collected for each data bit.

\begin{figure}[!t]
\centering
\includegraphics[width=\columnwidth]{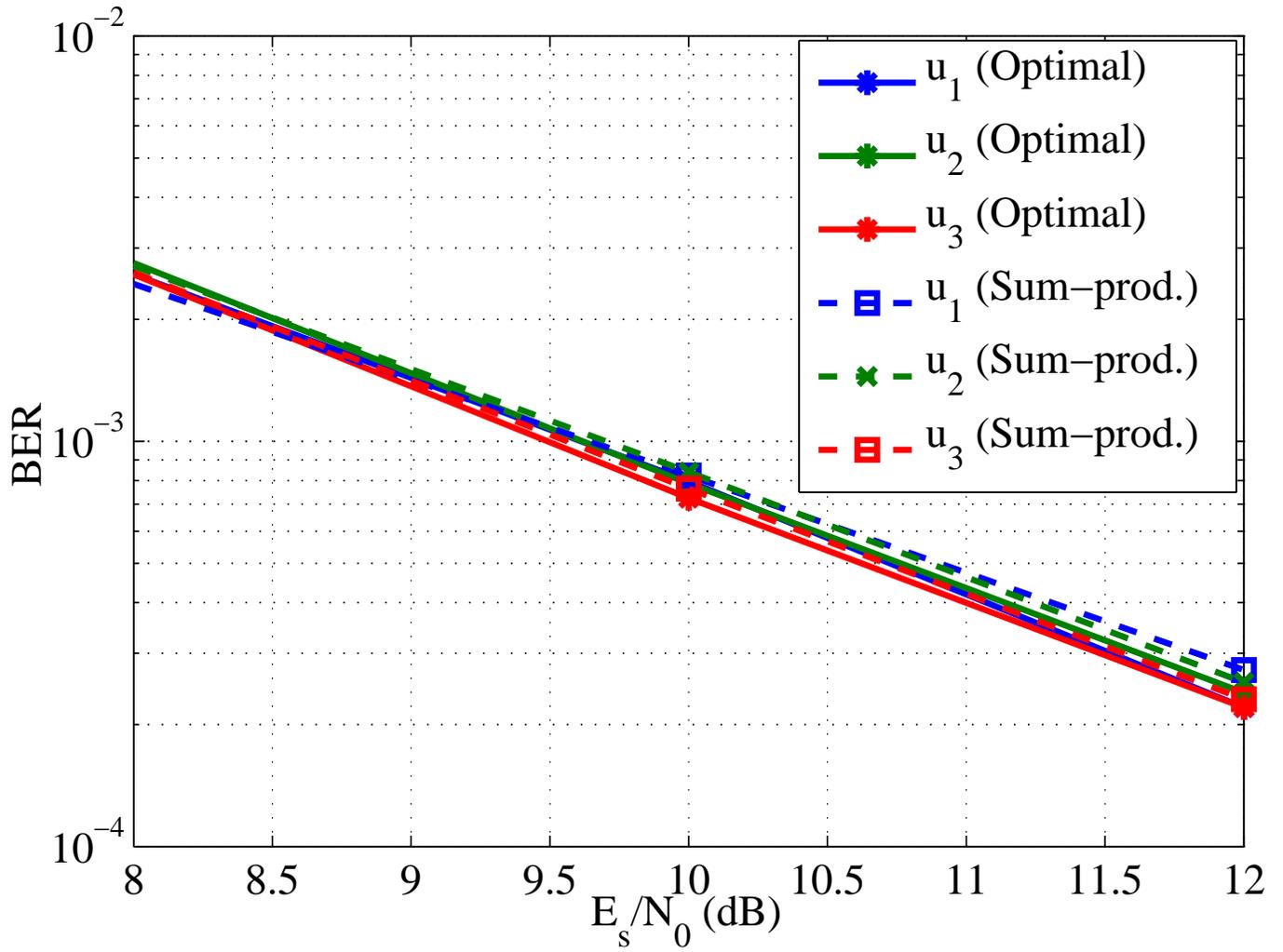}
\caption{BER curves for the individual MAP decoder of (\ref{eqn:opt_det}) and the SP iterative decoder}
\label{fig:ber_map_sum}
\end{figure}

In Fig.~\ref{fig:ber_map_sum}, we identify the fact that the SP decoder maintains almost the same BER performance as the optimal decoding rule. The SNR loss due to usage of the SP decoder is less than $0.1$ dB for a BER value of $10^{-3}$ for all data bits. Achieving full-diversity with a linear complexity order, SP type decoding may serve as an ideal method for decoding for the network coded system of (\ref{eqn:gen3}) despite the fact that the corresponding Tanner graph contains cycles. Results demonstrating the good performance of SP decoding were also reported previously in \cite{Aktas08, Moon04, Colavolpe2005} for graphs with cycles. In fact, one may realize that the length of the shortest cycle in the corresponding graph is $6$, hence the graph is said to have a girth of $6$. In \cite{Colavolpe2005} within the context of sparse intersymbol interference (ISI) channels, it is shown that for any graph with girth $6$, the performance of the SP algorithm is practically optimal. On the other hand, one may identify that for the family of greedy codes for $k=3$ users with blocklenghts larger than $6$, the girth of the corresponding graph will always be $4$. Fortunately, it is also given in \cite{Colavolpe2005} that the method of \textit{stretching} on girth-$4$ graphs yields modified girth-$6$ graphs on which the SP algorithm evaluates the APPs for the data symbols with negligible performance loss. Further details on the girth profile and degree distribution optimization procedures (like in \cite{Bocharova2012} for a greedy search of LDPC codes and like in \cite{Boutros2007} for root-check LDPC code design) are out of scope of this work.

\subsection{Performance of Selective Network Coding (SNC)}
\label{sec:perf_snc}
\begin{figure}[!t]
\centering
\includegraphics[width=\columnwidth]{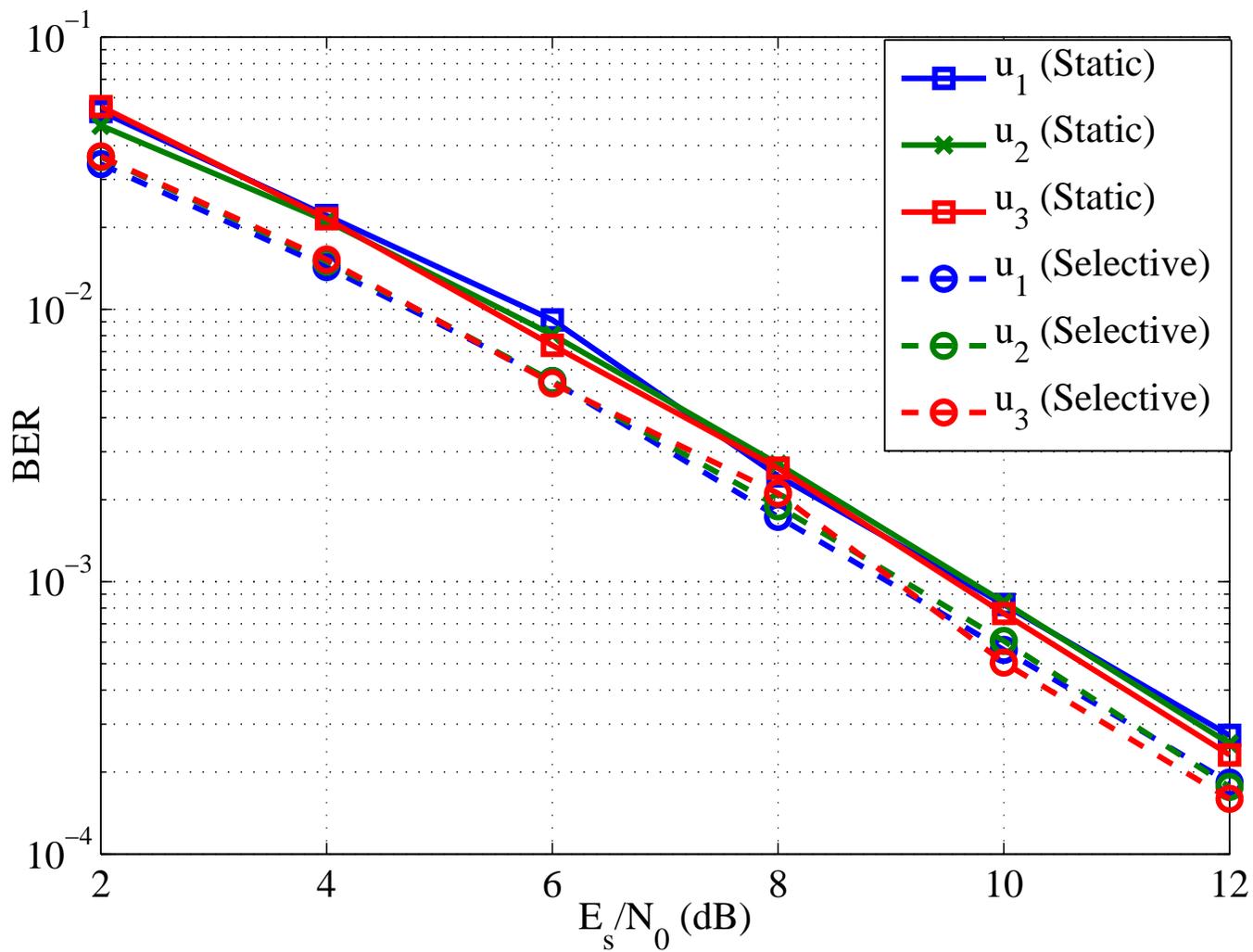}
\caption{Selective and static network encoding BER curves}
\label{fig:ber_selective_direct}
\end{figure}

The selective network encoding operation defined in Section \ref{sec:snc} is applied in this section on the Sample Network-II of Section \ref{sec:num_results_sub2}. The performance improvement for the selective encoding over the static (using fixed $\mathbf{G}$ with no selection of symbols to be encoded) encoding method is again shown using the SP iterative decoder of Section \ref{sec:spnd}. The instantaneous intermediate node error probabilities are compared with average error probabilities (dictated by $\mathbf{G}$) and data of the nodes whose error probabilities are below the corresponding average values (thresholds) are combined by the intermediate node. In Fig.~\ref{fig:ber_selective_direct}, we observe that SNC offers an SNR improvement  of $0.6$ dB for BER set to $10^{-3}$ over the static NC method. 
\subsection{Performance of NC under Slow-Fading Channel Model}
\label{sec:perf_slow}
\begin{figure}[!t]
\centering
\includegraphics[width=\columnwidth]{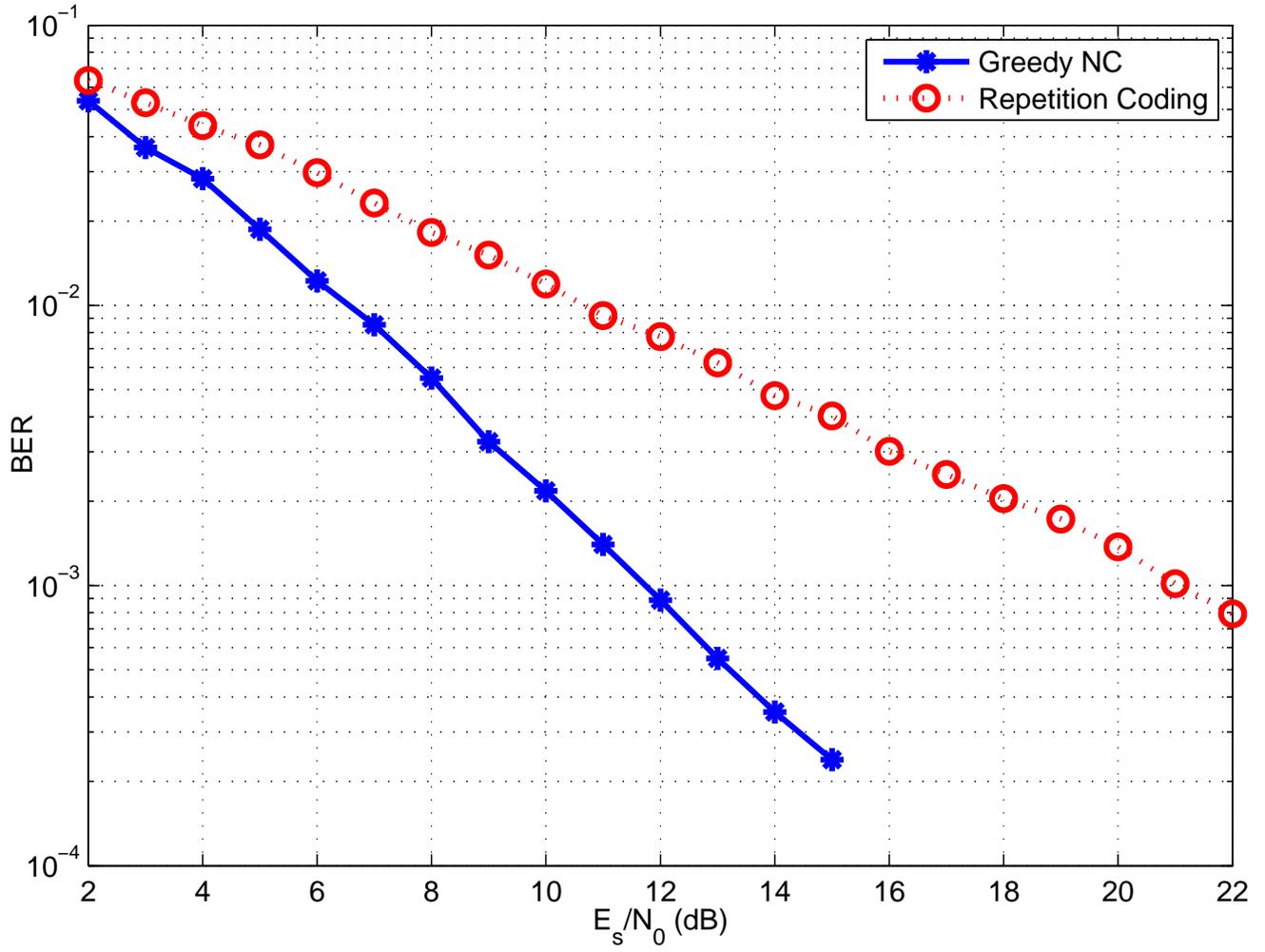}
\caption{BER vs. SNR curves for slow fading channel}
\label{fig:slow_fading}
\end{figure}

All the discussion and the results presented up until this section rely on the assumption that all channel gain coefficients related to the observations at node $0$ are independent. Hence a block fading model over time slots is utilized. However, it is also possible under many communication scenarios that the variation of a channel gain coefficient is not rapid enough for such an assumption. Then it is also possible that all transmissions from a selected source node to node $0$ observe the same fading condition leading to the degradation in BER performance due to loss in diversity. Therefore, we finalize the numerical results by providing the BER curves of NC and repetition coding under the assumption that within a round of $n$ transmissions, only the transmissions from distinct source nodes observe independent fading, i.e., $h_j$ and $h_m$ are independent if $v_j \neq v_m$ and otherwise $h_j=h_m$. In Fig. \ref{fig:slow_fading}, we investigate the BER curves for the Sample Network-II operating under this slower fading assumption. The repetition coding is represented by (\ref{eqn:gen2}) and NC is realized by (\ref{eqn:gen3}). It is seen that the repetition coding merely results in a diversity order of $1$ for each symbol as expected. On the other hand, NC yields an order of $2$ via the cooperative diversity obtained due to intermediate nodes transmitting over independent channels. The SNR losses incurred by not utilizing NC are shown to further increase in great amounts for this slower fading channel scenario.
 
\section{Conclusions}
\label{sec:conc} 
We formulated a NC problem for cooperative unicast transmissions. A generator matrix $\mathbf{G}$ and a scheduling vector $\mathbf{v}$ are used to represent the linear combinations performed at intermediate nodes. We presented a MAP-based decoding rule utilizing $\mathbf{G}$, $\mathbf{v}$, and the error probabilities at the intermediate nodes. A method for obtaining the performance determining parameter as the diversity order for individual source nodes is proposed for any given $\mathbf{G}$ over the corresponding separation vector. Through simulations we showed that our decoding rule, using reliability information for the network coded symbols, avoids the diversity order losses due to the error propagation effect. We presented design examples for network codes via greedy block codes, which may also provide unequal diversity orders to nodes with proper puncturing. Over given design examples, we obtained the rate-diversity trade-off curves and the rate advantage realized by using NC with respect to the no NC case. Moreover, the SP iterative network decoder with linear complexity order is proposed and shown to perform quite close to the optimal rule. Furthermore, the selective NC scheme combining only the reliably detected data at cooperating nodes is shown to yield additional coding gains. Identifying gains of NC for purely random $\mathbf{G}$ matrices in large networks, studying the effects of imperfect information on channel gains and relay error probabilities will be addressed in future work. Finally, it would be also interesting to operate suggested wireless NC methods under asymmetrical channel gains, which can be more realistic for ad hoc networks.

\bibliography{Practical_Methods_for_Wireless_Network_Coding_with_Multiple_Unicast_Transmissions_Tugcan_Aktas}
\begin{figure}[htbp]
\centering
\includegraphics[width=\columnwidth]{figures/network3}
\caption{Sample network coded transmission scenario} 
\label{fig:network3}
\end{figure}
\end{document}